\documentclass[onecolumn,showpacs,floatfix,11pt,nofootinbib]{revtex4}

\def\ben{\begin{eqnarray}}
\def\een{\end{eqnarray}}
\def\bq{\begin{quote}}
\def\eq{\end{quote}}

 at 10truept

\newcommand{\beq}{\begin{equation}}
\newcommand{\eeq}{\end{equation}}
\newcommand{\beqa}{\begin{eqnarray}}
\newcommand{\eeqa}{\end{eqnarray}}

\usepackage{amssymb,amsmath,wasysym}
\usepackage{graphicx,float}
\usepackage{indentfirst}
\newcommand{\be}{\begin{equation}}
\newcommand{\ee}{\end{equation}}

\newcommand{\ba}{\begin{eqnarray}}
\newcommand{\ea}{\end{eqnarray}}

\begin{document}\title{Spherically Symmetric Thick Branes Cosmological Evolution}
\author{A. E. Bernardini}
\email{alexeb@ufscar.br}
\affiliation{Departamento de F\'{\i}sica, Universidade Federal de S\~ao Carlos, PO Box 676, 13565-905, S\~ao Carlos, SP, Brazil}
\author{R. T. Cavalcanti}
\email{rogerio.cavalcanti@ufabc.edu.br}
 \affiliation{
Centro de Ci\^encias Naturais e Humanas,
Universidade Federal do ABC, 09210-580, Santo Andr\'e, SP, Brazil}
\author{Rold\~ao da Rocha}
\email{roldao.rocha@ufabc.edu.br}
\affiliation{Centro de Matem\'atica, Computa\c c\~ao e Cogni\c c\~ao, Universidade Federal do ABC,  09210-580, Santo Andr\'e, SP, Brazil}
\affiliation{International School for Advanced Studies (SISSA), Via Bonomea 265, 34136 Trieste, Italy.}


\begin{abstract}
Spherically symmetric time-dependent solutions for the 5D system of a scalar field canonically coupled to gravity are obtained and identified as an extension of recent results obtained by Ahmed, Grzadkowskia and Wudkab \cite{aqeel}.
The corresponding cosmology of models with regularized branes generated by such a 5D scalar field scenario is also investigated.
It has been shown that the anisotropic evolution of the warp factor and consequently the Hubble like parameter are both driven by the radial coordinate on the brane, which leads to an emergent thick brane-world scenario with spherically symmetric time dependent warp factor.
Meanwhile, the separability of variables depending on fifth dimension, $y$, which is exhibited by the equations of motion, allows one to recover the extra dimensional profiles obtained in Ref.~\cite{aqeel}, namely the extra dimensional part of the scale (warp) factor and the scalar field dependence on $y$.
Therefore, our results are mainly concerned with the  time dependence of a spherically symmetric warp factor.
Besides evincing possibilities for obtaining asymmetric stable brane-world scenarios, the extra dimensional profiles here obtained can also be reduced to those ones investigated in \cite{aqeel}.
\end{abstract}


\date{\today}
\maketitle
\newpage
 
Brane-world models are a straightforward 5D phenomenological realization of the Ho$\check{\rm r}$ava-Witten supergravity solutions~\cite{HW}, where the hidden brane is placed at infinity and the {\em moduli} effects from compact extra dimensions are neglected \cite{maartens}. 
Once introduced in the context of an effective theory of supergravity on domain walls \cite{Rey}, brane-world scenarios
are supported by seminal results \cite{RS,RS1,PREV,prev1} which are relevant in realizing 4D gravity on a domain wall in 5D space-time \cite{RS,RS1,Cvetic01}. 

Brane-world cosmology has also been investigated in several suitable contexts.
Classes of exact solutions with a constant 5D radius on a cosmologically evolving brane were provided in \cite{Binetruy1}, allowing unconventional cosmological equations with the matter content of the brane dominating that of the bulk.
This framework is in full compliance to standard cosmology, as the present values of the Hubble parameter and of the cosmological background radiation temperature fits their respective values at the time of nucleosynthesis.
Moreover, brane-world cosmology in thin branes has been studied for any equation of state describing the matter in the brane, where standard cosmological evolution can be obtained after an early non-conventional phase in typical Randall-Sundrum \cite{Binetruy2} scenarios, where the brane tension compensates the bulk cosmological constant. 
The accelerated Universe could be the result of the gravitational leakage into extra dimensions on Hubble distances rather than the consequence of non-zero cosmological constant \cite{Deffayet:2001pu}.
Some attempts of devising the Friedmann law on the brane have involved a dark radiation term due to the bulk Weyl tensor, which depends linearly on the brane energy densities.
For any equation of state on the brane, the radiation was shown to evolve such as to generate conventional radiation-dominated cosmology, consistent with nucleosynthesis \cite{Khoury:2002jq}.
 
Subsequent to the brane-world cosmology on thin branes, the thick brane-world paradigm has exhibited a fine structure \cite{aq1,aqeel, Ahmed2} that supports the above discussed phenomenology.
In spite of their success, thick brane-world models do encompass neither anisotropy on the brane nor the important framework of asymmetric branes, as well as spherically symmetric thick brane worlds. 
 
Even if anisotropic brane-worlds have been comprehensively investigated, there still are several reasons to depart from the standard isotropic models, in particular focusing on spherically symmetric time-dependent warp factors, in the thick brane-world scenario. 
As an example, in Bianchi I brane-world cosmology, for scalar fields with a large kinetic term, the initial expansion of the Universe is quasi-isotropic.
The Universe grows anisotropically during an intermediate transient regime and the anisotropy finally disappears during the inflationary expansion \cite{maartensani}.
In addition, anisotropic brane-worlds are realized in the context of exact solutions of the gravitational field equations in the generalized Randall-Sundrum model for an anisotropic brane with Bianchi type I and V geometry, with perfect fluid and scalar fields as matter sources.
Under the assumption of a conformally flat bulk with vanishing Weyl tensor for a cosmological fluid obeying a linear barotropic equation of state, the general solution of the field equations was expressed as an exact parametric form for both Bianchi type I and V spacetimes \cite{harko}.
The dynamics of the corresponding anisotropy in such Bianchi type I and V cosmological scenarios has also been investigated in the context of Randall-Sundrum brane-worlds \cite{26}, given that a Randall-Sundrum brane-world can be mimicked without the assumption of spatial isotropy, by means of an homogeneous and anisotropic Kasner type solution of the Einstein-AdS equations in the bulk \cite{28}.
Other interesting anisotropic brane-world models have been studied in Refs.
~\cite{frolov,olps2013}.

On the other hand, observations of the CMB tell one that the Universe is isotropic with a great accuracy \cite{52}.
The natural framework to approach this highly isotropic Universe implies into assuming that the Universe setup from a highly anisotropic state, and thus a dynamical mechanism gets rid of almost all its anisotropy.
Inflation mechanisms \cite{54, 55} are the most promising candidates for explaining such a behavior.
In these lines, the simplest generalization of FRW cosmologies are the Bianchi cosmologies, as they provide anisotropic but homogeneous cosmologies, where the central point of discussion is if the Universe can isotropize without additionally fine-tuning the parameters of the model.
The isotropization of Bianchi I brane-world cosmologies has been investigated, from several points of view \cite{20,58}.
It has been shown, for instance, that a large initial anisotropy does not suppress inflation in a Bianchi I brane-world \cite{maartensani}. 
Otherwise, considering negative values of dark radiation in Bianchi I models leads to interesting solutions for which the Universe can both collapse or isotropize \cite{26,18,20}.
It is required that isotropization should be accompanied by a phase of accelerated expansion in order to be a good candidate to explain the results that indicate the current speeding up of the observable Universe \cite{61}.
This latter observational fact is approached from two directions: modifying the gravitational sector \cite{34} or introducing dark energy \cite{63}.
From this point of view, a model in which a dark energy component lives in a Bianchi brane-world combines both approaches. 

The study of the dynamics of a scalar field with an arbitrary potential trapped in brane-world model can be further performed \cite{aqeel,Ahmed2,soda,wang}.
Homogeneous and anisotropic Bianchi I branes filled also with a perfect fluid are the mostly approached models.
In particular, by taking into account the effect of a positive dark radiation term on the brane \cite{Escobar:2013js}, the effect of the projection of the 5D Weyl tensor onto the brane in the form of a negative dark radiation term is considered \cite{23}. 

All the above-mentioned reasons motivate the investigation of both spherically symmetric and anisotropic brane models.
It is indeed worthwhile to emphasize that cosmological solutions of the gravitational field equations in the generalized Randall-Sundrum model for an anisotropic brane were obtained, with Bianchi I geometry and with perfect fluid as matter sources described by a scalar field \cite{Bernardini12A}.
The solution admits an inflationary era and, at a later epoch, the anisotropy of the Universe washes out.
Two classes of cosmological scenarios are involved, regarding universes that evolve from a singularity and without singularity \cite{Paul:2001ce}. 
Moreover, by using a metric-based formalism to treat cosmological perturbations \cite{Alex01,Alex02}, the connection between anisotropic stress on the brane and brane bending are discussed in \cite{Dorca:2000iq}.

Our main aim here is to provide  an initial approach for spherically symmetric thick brane cosmology.
By exploring the framework of isotropic thick branes \cite{aq1,aqeel, Ahmed2}, one can realize that the separability of the warp factor is fundamental in order 
to explicitly describe the time-dependent solutions.
It is noway obvious that, for spherically symmetric thick brane-worlds, the  warp factors to be considered in this paper -- that are dependent on time, extra dimension, and radial coordinate on the brane -- should be separable in the context of solving the equations of motion. 
Likewise, it suggests that it might be hopeful to find time-dependent soliton solutions leading to non-separable forms of the warp factor \cite{g1,g2,g3}.
Separable solutions are normally discussed in the framework of thin brane-world models that are rather unnatural in case of thick defects, since the brane thickness $\Delta$ must fulfill  the limits $2.0\times 10^{-19} {\rm m}\lesssim \Delta\lesssim 44\mu{\rm m}$ \cite{Kapner:2006si}, having thus a minimal thickness \cite{HoffdaSilva:2012em}.
In fact, thick brane cosmology has been widely 
discussed in \cite{g1,g2,g3,German:2012rv}, further regarding other type of warp factors \cite{gog1,gog2,Carrillo-Gonzalez:2013exa,Cruz:2011kj,Bernardini13} and tachyonic solutions, with a decaying warp factor that enables localization of 4D gravity as well as other matter fields \cite{gog3}. Some applications in the thin brane limit have been provided, e. g., in \cite{daRocha:2012pt,CoimbraAraujo:2005es,HoffdaSilva:2009ht}.

Departing from a general 5D spherically symmetric warped spacetime, our purpose is to solve the coupled system of Einstein equations and the equations of motion for a scalar field.
The procedure introduced in the following results into an explicit formula for both the extra dimension-dependent part of the warp factor and the spherically symmetric time-dependent component.
The warp factors for flat, closed, and open spacetimes are obtained and discussed, and the properties of Hubble type parameter are also investigated.
Our analysis results into deploying the fundamentals of thick brane-world cosmology with time dependent spherically symmetric warp factors, exclusively departing from the Einstein equations. 

 
To provide a generalization of the successful achievements on brane-world cosmology in the thin brane paradigm \cite{Binetruy1, Binetruy2} as well as in the thick brane scenario \cite{aqeel,Ahmed2}, one considers 5D spacetimes for which the metric assumes
the following form:
\begin{align}
ds^2&=a^2(t,r,y)g_{\mu\nu}dx^\mu dx^\nu+dy^2, \label{metric}
\end{align}
where $x^\mu$ denotes a chart of 4D coordinates on the brane, whereas $g_{\mu\nu}$ is the metric given by 
\begin{eqnarray}
g_{\mu\nu}dx^\mu dx^\nu = -dt^2 + \left( \frac{dr^2}{1-kr^2} + r^2d\Omega_2\right), \nonumber
\end{eqnarray}
where $d\Omega_2$ stands for the usual area element of the 2-sphere and $k$ denotes the curvature parameter assuming the values $-1$, 0 and $1$, leading respectively to an open, a flat or a closed Universe. 
The function $a(t,r,y)$ is the conformal scale factor extraordinarily depending upon the radial coordinate $r$ on the brane, also referred as a warp factor due to the extra dimension $y$ in (\ref{metric}). The 4D solutions are sourced by the bulk scalar field. 

The action for scalar field in the presence of 5D gravity is given by 
\begin{equation}
S=\int d^5x \sqrt{-{\rm g}}\left(-\frac{1}{2}{\rm g}^{MN}\nabla_{M}\phi\nabla_{N}\phi-V(\phi)+2M_5^{3}R\right),
\label{action}
\end{equation}
where g denotes the 5D metric, $M_5$ is the Planck mass of the fundamental 5D theory and $R$ denotes the 5D Ricci scalar.
For the above prescribed scenario, one assumes that the scalar field, $\phi$, depends exclusively on time and upon the extra coordinate, $y$, and $V(\phi)$ is the scalar field potential.

The Einstein equations and the equation of motion for $\phi$ resulting from the above action \eqref{action} are provided by 
\begin{eqnarray}
\nabla^2\phi-\frac{dV}{d\phi}&=&0, \label{eineq2}\\
R_{MN}-\frac{1}{2}{\rm g}_{MN}R&=&\frac{1}{4M_5^{3}}T_{MN},\label{eineq1}
\end{eqnarray}
where $\nabla^2$ is the 5D Laplacian operator, and the energy-momentum tensor, $T_{MN}$, for the
scalar field $\phi(t,y)$ reads
\begin{equation}
T_{MN}=-{\rm g}_{MN}\left(\frac{1}{2}(\nabla\phi)^{2}+V(\phi)\right)+\nabla_{M}\phi\nabla_{N}\phi\,.
\nonumber\label{emt}
\end{equation}
In particular, the energy-density ($T_{00}$) is implied by $\phi(y)$ and localized near $y = 0$. Moreover, the  equation of motion for the scalar field is expressed by
\beq
\phi''-\frac{1}{a^2}\ddot\phi+\frac{4{a'}}{a}\phi' -\frac{2\dot a}{a^3}\dot\phi=\frac{dV}{d\phi}\,, \nonumber
\eeq
where one denotes 
$\frac{\partial f}{\partial t} = \dot{f},\; \frac{\partial f}{\partial r} = \bar{f},$ and $ \frac{\partial f}{\partial y} = {f'}
$, for any scalar function $f$ hereupon. 
By assuming a static scalar field scenario, the components of the Einstein tensor
are given by the following expressions:
\begin{eqnarray}\nonumber
G_{00}&=&a^{2}\left\{3\left[ \frac{\dot{a}^{2}}{a^{4}}-\left( \frac{a''}{a}+\frac{a'^{2}}{a^{2}}\right)+\frac{k}{a^{2}}\right] +(1-kr^{2})\left[\frac{\bar{a}^{2}}{a^{4}}-\frac{2\bar{\bar{a}}}{a^{3}}-\frac{6\bar{a}}{a^{3}r} \right]+\frac{2\bar{a}}{a^{3}r} \right\},\\\nonumber 
G_{11}&=&-g_{11}\left\{\left[ \frac{1}{a^{2}}\left(\frac{2\ddot{a}}{a}-\frac{\dot{a}^{2}}{a^{2}}\right)-3\left( \frac{a''}{a}+\frac{a'^{2}}{a^{2}}\right)+\frac{k}{a^{2}}\right] -(1-kr^{2})\left[\frac{3\bar{a}^{2}}{a^{4}}+\frac{4\bar{a}}{a^{3}r} \right] \right\},\\\nonumber
G_{22}&=&-g_{22}\left\{\left[ \frac{1}{a^{2}}\left(\frac{2\ddot{a}}{a}-\frac{\dot{a}^{2}}{a^{2}}\right)-3\left( \frac{a''}{a}+\frac{a'^{2}}{a^{2}}\right)+\frac{k}{a^{2}}\right] -(1-kr^{2})\left[\frac{2\bar{\bar{a}}}{a^{3}}-\frac{\bar{a}^{2}}{a^{4}}+\frac{4\bar{a}}{a^{3}r} \right]+\frac{2\bar{a}}{a^{3}r} \right\},\\ \nonumber
G_{33}&=&{g_{33}}G_{22}/{g_{22}},\\ \nonumber
G_{55}&=&3\left(\frac{2a'^{2}}{a^{2}}- \frac{\ddot{a}}{a^3}-\frac{k}{a^{2}}\right)+3(1-kr^{2})\left(\frac{\bar{\bar{a}}}{a^{3}}+\frac{3\bar{a}}{a^{3}r} \right)-\frac{3\bar{a}}{a^{3}r}, \\ \nonumber
G_{01}&=&2\left(2\frac{\bar{a}\dot{a}}{a^{2}}-\frac{\dot{\bar{a}}}{a} \right),\\ \nonumber
G_{05}&=&3\left(\frac{\dot{a}{a'}}{a^{2}}-\frac{\dot{{a}}'}{a} \right),\\ \nonumber
G_{15}&=&3\left(\frac{\bar{a}{a'}}{a^{2}}-\frac{{\bar{a}}'}{a} \right).
\end{eqnarray}

The Einstein equations $G_{MN}=T_{MN}$, when $4M_*^{3} = 1$, can be used to find the 
form of the warp factor. 
By separating the variables $a(t,r,y)=A(t,r)B(y)$, the Einstein equation $G_{01}=T_{01}=0$ yields 
\begin{eqnarray}
0&=&
2\frac{\bar{A}}{A}-\frac{\dot{\bar{A}}}{\dot{A}}=\partial_r\ln A^{2}- \partial_r \ln \dot{A}\nonumber\\
&\Rightarrow & \ln \frac{A^{2}}{\dot{A}}=T(t) \Leftrightarrow \dot{A}=A^{2}e^{-T}\,,\label{dota}\end{eqnarray}
or
\begin{eqnarray}
0&=&2\frac{\dot{A}}{A}-\frac{\dot{\bar{A}}}{\bar{A}}=\partial_t\ln \frac{A^{2}}{\bar{A}}\nonumber\\\label{bara}
&\Rightarrow & \ln \frac{A^{2}}{\bar{A}}=R(r) \Leftrightarrow \bar{A}=A^{2}e^{-R}\,,
\end{eqnarray}
implying that 
\begin{eqnarray}
 \label{dotbar}
 \dot{A}=\bar{A}e^{R-T}\,.
\end{eqnarray}
The expressions $\dot{A}^{2}=A^{4}e^{-2T}$ and $\bar{A}^{2}=A^{4}e^{-2R}$ follow from \eqref{dota}, \eqref{bara} and \eqref{dotbar}, and they imply that 
\begin{eqnarray}
\dot{\bar{A}}&=&
=2A^{3}e^{-(T+R)}\,,\nonumber\\
\ddot{A}&=&
=A^{2}e^{-T}\left(2Ae^{-T}-\dot{T}\right)\,,\label{aa0}\\
\bar{\bar{A}}&=&
A^{2}e^{-R}\left(2Ae^{-R}-\bar{R}\right)\,.\label{aa1}
\end{eqnarray}
One of the off-diagonal Einstein equations $G_{05}=T_{05}=\dot{\phi}\phi'$ yields
\begin{eqnarray}
\dot{\phi}\phi'&=&3
\left(\frac{\dot{A}B'}{AB}-\frac{\dot{A}B'}{AB} \right)=0 \;\;\;\Rightarrow\;\;\; \dot{\phi}=0\,,\nonumber
\end{eqnarray}
that means that the scalar field $\phi$ is time independent.

Hence the components of energy-momentum tensor become
\begin{eqnarray}
T_{\mu \nu}&=&-g_{\mu \nu}\left(\frac{1}{2}\phi'^{2} + V(\phi) \right)\,, \qquad\quad
T_{55}=\frac{1}{2}{\phi'}^{2} - V(\phi)\,. 
\nonumber\end{eqnarray}
The explicit form of the components of the Einstein equation for the metric ansatz \eqref{metric}
can be written for $k = 0,~\pm1$, by 
denoting the spatial curvature of the 4D homogeneous and isotropic
space-time for Minkowski, de Sitter and anti-de Sitter space, respectively.
The diagonal components (remembering that the component $_{22}$ equals $_{33}$) are respectively expressed as:
\begin{eqnarray}\nonumber
\frac{1}{2}\phi'\!+\!V(\phi) &=&\frac{1}{B^{2}}\!\left\{\! -3\left(B{B'}\right)' \!+\!\frac{1}{A^{2}}\left[(1-kr^{2})\left(\frac{\bar{A}^{2}}{A^{2}}-4\frac{\bar{\bar{A}}}{A}-6\frac{\bar{A}}{Ar} \right)\!+\!2\frac{\bar{A}}{Ar}\!+\!3\frac{\dot{A}^{2}}{A^{2}}\!+\!3{k}\right]\right\},\nonumber\\
\frac{1}{2}\phi'\!+\!V(\phi) &=&\frac{1}{B^{2}}\!\left\{\! -3\left(B{B'}\right)' \!+\!\frac{1}{A^{2}}\left[(1-kr^{2})\left(3\frac{\bar{A}^{2}}{A^{2}}\!+\!4\frac{\bar{A}}{Ar} \right)\!+\!2\frac{\ddot{A}}{A}-\frac{\dot{A}^{2}}{A^{2}}\!+\!{k}\right]\right\},\nonumber\\
\frac{1}{2}\phi'\!+\!V(\phi) &=&\frac{1}{B^{2}}\!\left\{\! -3\left(B{B'}\right)' \!+\!\frac{1}{A^{2}}\left[(1-kr^{2})\left(2\frac{\bar{\bar{A}}}{A}-\frac{\bar{A}^{2}}{A^{2}}\!+\!4\frac{\bar{A}}{Ar} \right)\!+\!2\frac{\ddot{A}}{A}-\frac{\dot{A}^{2}}{A^{2}}\!+\!{k}\!+\!2\frac{\bar{A}}{Ar}\right]\right\},\nonumber\\
\frac{1}{2}\phi'-V(\phi)
 &=&-\frac{1}{B^{2}}\left\{ -6{B'^{2}} \!+\!\frac{3}{A^{2}}\left[(kr^{2}-1)\left(\frac{\bar{\bar{A}}}{A}\!+\!3\frac{\bar{A}}{Ar} \right)\!+\!\frac{\ddot{A}}{A} \!+\!{k}\!+\!\frac{\bar{A}}{Ar} \right]\right\}.\label{by}
 \end{eqnarray}
 Therefore, the equations for $\phi$ or $y$
 can be expressed as
\begin{eqnarray}
\phi''(y)+4\frac{{a'}}{a}{\phi'}(y)&=&\frac{dV}{d\phi},\nonumber\\B^2(y)\left[ 3\left( \frac{B''(y)}{B(y)}+{\frac{B'^2(y)}{B^2(y)}}\right)+\frac{1}{2}\phi'(y)+V(\phi)\right]&=&c_0,\nonumber\\
B^2(y)\left[ 6{\frac{B'^{2}(y)}{B^2(y)}}-\frac{1}{2}\phi'(y)+V(\phi)\right]&=&c_5\,,\label{c0c5}\end{eqnarray} for $c_0$ and $c_5$ separation constants, whereas the ones for $t$ and $r$ are summarized 
respectively by
\begin{eqnarray}
&&\dot{A}(t,r)=A^2(t,r)e^{-T(t)},\nonumber\\
&&\bar{A}(t,r)=A^2(t,r)e^{-R(r)},\label{a2a}\\\label{e00}
_{00}:&& \frac{1}{A^{2}}\left[(1-kr^{2})\left(\frac{\bar{A}^{2}}{A^{2}}-\frac{4\bar{\bar{A}}}{A}-\frac{6\bar{A}}{Ar} \right)+2\frac{\bar{A}}{Ar}\right]+\frac{3\dot{A}^2}{A^{4}}+\frac{3k}{A^2}=c_0,\\
\label{e11}
_{11}: &&\frac{1}{A^{2}}\left[(1-kr^{2})\left(\frac{3\bar{A}^{2}}{A^{2}}+\frac{4\bar{A}}{Ar} \right)\right]+\frac{2\ddot{A}}{A^3}-\left(\frac{\dot{A}}{A^{2}}\right)^2+\frac{k}{A^2}=c_0,\\\label{e22}
_{22}\, =\; _{33}: &&\frac{1}{A^{2}}\left[(1-kr^{2})\left(\frac{2\bar{\bar{A}}}{A}-\frac{\bar{A}^{2}}{A^{2}}+\frac{4\bar{A}}{Ar} \right)+\frac{2\bar{A}}{Ar}\right]+\frac{2\ddot{A}}{A^3}-\frac{\dot{A}^2}{A^{4}}+\frac{k}{A^2}=c_0,\\\label{e55}
_{55}: &&\frac{3}{A^{2}}\left[(kr^{2}-1)\left(\frac{\bar{\bar{A}}}{A}+\frac{3\bar{A}}{Ar} \right)+\frac{\bar{A}}{Ar}\right]+\frac{3\ddot{A}}{A^3} +\frac{3k}{A^2} =c_5.
\end{eqnarray}
 The role of the bulk scalar field is to provide the cosmological constant on a brane as is clear from Eqs.(\ref{a2a}-\ref{e22}). 
By imposing $c_5 = \Lambda$ one has analogous cosmological implications for suitable limits, where the warp factor has no dependence on $r$, as in the thick brane cosmology with isotropic warp factor \cite{aqeel,Ahmed2}. In an isotropic thick brane-world the condition $c_5 = 2c_0=\Lambda$ holds \cite{aqeel}. Nevertheless, in this scenario 
such two constants restrict further the form of the function $A(r,t)$, when the above equations are used, by the following 
 relationship:
\begin{eqnarray}
-15\left(\frac{kr^2-1}{c_1r}\right)^2-12\frac{(1-kr^2)^{3/2}}{c_1Ar^2}+6\frac{\sqrt{1-kr^2}}{c_1Ar^2}&=&2c_0 - c_5\,.\label{opop}
\end{eqnarray}
Note that this consistency equation is trivial if the 4D scale factor is independent of the radial coordinate, as in \cite{aqeel}.
Now, by computing the difference of \eqref{e11} and \eqref{e22}, one obtains the equation 
$
(1-kr^{2})\bar{R}-\frac{1}{r}=0\nonumber
$ which has solution
\begin{equation}
R(r)=\ln \frac{c_1r}{\sqrt{1-kr^{2}}}\,, 
\end{equation} where $c_1$ is a constant 
of integration. 
Moreover, the solution for Eq.~(\ref{a2a}) 
is provided by (hereon one shall notice the index $k$ in order to denote the dependence on $k=0,\pm 1$ in the following expressions):
\begin{equation}\label{A1}
A_k(t,r)=\frac{c_1}{c_1Y_k(t)+f_k(r)}\,,
\end{equation}
\noindent where $Y_k(t)$ is a constant of integration with respect to the $r$ coordinate,
\begin{eqnarray}\label{fk}
f_k(r)&\equiv&\ln \frac{\sqrt{1-kr^{2}}+1}{r}-\sqrt{1-kr^{2}}\,,
\end{eqnarray} and $A_k(t,r)$ depends on $k=0,\pm 1$. It implies hence that \begin{equation}\label{aa2}
\frac{\dot{A}_k(t,r)}{A^2_k(t,r)}=-\dot{Y}_k(t)=e^{-T(t)}\,. 
\end{equation}

We can simplify Einstein equations using (\ref{bara}), (\ref{aa0}), (\ref{aa1}), and (\ref{aa2}) to make 
 Eq.\eqref{e00} --- that corresponds to the $_{00}$ component of the Einstein equations --- to read: 
\begin{equation}\label{eqY}
\dot{Y}_k^2=-kY_k^2+w_k(r)Y_k+z_k(r)\,,
\end{equation}
where $w_k(r)$ and $z_k(r)$ are respectively given by the following expressions:
\begin{eqnarray} \nonumber
w_k(r)&=&-\frac{1}{3}\left[2(1-kr^{2})e^{-R}\left(2\bar{R}-\frac{3}{r} \right)+2\frac{e^{-R}}{r}+\frac{6kf_k(r)}{c_1}\right]\,,\\ \nonumber
z_k(r)&=&-\frac{1}{3}\left\{(1-kr^{2})e^{-R}\left[-7e^{-R}+\frac{2f_k(r)}{c_1}\left(2\bar{R}-\frac{3}{r} \right) \right]+\frac{2f_k(r)}{c_1}\frac{e^{-R}}{r}+3k\frac{f_k^2(r)}{c_1^2} -c_0\right\}\,.
\end{eqnarray}
 
The solutions for Minkowski, anti-de Sitter and de Sitter spacetimes are respectively provided by:\begin{eqnarray}\label{y1}
Y^{\pm}_{-1}(t)&=&\frac{1}{4}e^{\pm(t\mp \alpha_{-1})}\left[\left( e^{\mp(t\mp \alpha_{-1})} -w_{-1}(r) \right)^2-4z_{-1}(r) \right],\\
\label{y0}
Y^{\pm}_0(t)&=&\alpha_0 \pm \sqrt{z_0(r)}t,\\
\label{ym1}
Y^{\pm}_{1}(t)&=&\frac{1}{2}\left[w_{1}(r) \pm\sqrt{w_{1}^2(r)+4z_{1}(r)}\sin(t+|\alpha_1|) \right]\,.
\end{eqnarray}
respectively for $k=-1, 0, 1$. The constant parameters $\alpha_0, \alpha_{\pm 1}$ are constants of integration.


When $f_k(r)=0$, then $A_k(t,r)= 1/Y_k(t)$, and one has the results from \cite{aqeel} for thick brane cosmology, with $c_5 = 2c_0=\Lambda$, where $\Lambda$ denotes the 4D cosmological constant.
For the case explicitly provided by Eq.~(\ref{A1}), $A_k(t,r)$ indeed does not depend on the $r$ coordinate.
Firstly, it is evident that $A_k(t,r)=0$ when $r\rightarrow\infty$, as $f_k(r)$ diverges in this case.
However, it is not properly the useful case here.
For $k=0$, $A_k(t,r)$ is independent of $r$ when $f_k(r)=0$, namely, when $r = 2/e$.
Moreover, when $k = 1$, $f_k(r)=0$ for $r = 1$, and hence $A_k(t,r)$ in Eq.~(\ref{A1}) has no dependence on the variable $r$. Finally, $A_k(t,r)$ is merely a function of the cosmic time $t$ for $k = -1$ when $r$ solves the algebraic equation $\frac{\sqrt{1+r^2}+1}{r} = \exp\left(\sqrt{1+r^2}\right)$, which is $r_0\approx0.663$.

Eqs.~(\ref{y1})-(\ref{ym1}) lead to the solutions \cite{aqeel}
\begin{eqnarray}
{a}(t) &\sim&\begin{cases}
\sec{\!\rm h}(t+\alpha_{-1})\,, & k=-1 \quad (\Lambda<0)\\
1/(t+\alpha_0)\,, & k=0 \quad (\Lambda>0)\\
\sec(t+\alpha_1)\,,& k=+1 \quad (\Lambda>0)\end{cases}
\label{aqeel}
\end{eqnarray}
For $k = 0, 1$ the metric is singular at a finite time $t =-\alpha_a + (n + 1/2)\pi k$, ($a=0,1$), for $n\in\mathbb{Z}$ \cite{aqeel}.
It further implies that in this particular case the Hubble parameter reads
\begin{eqnarray}
{H}(t) &=&\begin{cases}
\tanh(t+\alpha_{-1})\,, & k=-1 \quad (\Lambda<0)\\
1/(t+\alpha_0)\,, & k=0 \quad (\Lambda>0)\\
\tan(t+\alpha_1)\,,& k=+1 \quad (\Lambda>0)\end{cases}
\label{aqeel1}
\end{eqnarray}
for the appropriate limits above analyzed where $f_k(r)=0$.

Once the $_{00}$ component of the Einstein equations is considered, one can further analyze the $_{11}$ component. Eq.~\eqref{e11} thus reads
\begin{eqnarray}\nonumber
&&(1-kr^{2})e^{-R}\left(3e^{-R}+\frac{4}{c_1r}f_k(r) \right)+\frac{k}{c_1²}f_k(r)²+kY\,\\
&&+\left[(1-kr^{2})e^{-R}\frac{4}{r} +\frac{2k}{c_1}f_k(r)\right]Y-2\frac{f_k(r)}{c_1}\ddot{Y}-2Y\ddot{Y}+3{\dot{Y}^{2}}=c_0\,.\nonumber
\end{eqnarray}
It implies that 
\begin{equation}\label{yh1}
2\ddot{Y}\left( Y+\frac{f_k(r)}{c_1}\right)=3\dot{Y}^2+kY^2+u(r)Y+v(r)\,,
\end{equation}
where $u_k(r)$ and $v_k(r)$ are respectively given by:
\begin{eqnarray} \nonumber
u_k(r)&=&\frac{4}{r}(1-kr^{2})e^{-R}+\frac{2k}{c_1}f_k(r)\,,\\ \nonumber
v_k(r)&=&u_k(r)\frac{f_k(r)}{c_1}+3(1-kr^{2})e^{-2R}-\frac{k}{c_1²}f_k(r)² -c_0\,.
\end{eqnarray}
Moreover, the $_{22}$ and $_{33}$ components of Einstein equations are provided by Eq.~\eqref{e22}, yielding
\begin{equation}\label{yh2}
2\ddot{Y}\left( Y+\frac{f_k(r)}{c_1}\right)=3\dot{Y}^2+kY^2+m(r)Y+n(r)\,,
\end{equation}
where
\begin{align}
m(r)&=e^{-R}\left[-2(kr^2-1)\bar{R}+\frac{6}{r}-4kr\right]+2k\frac{f_k(r)}{c_1}\,,\nonumber\\
n(r)&=m(r)\frac{f_k(r)}{c_1}+3e^{-2R}(1-kr^2)-k\frac{f_k^2(r)}{c_1^2}-c_0\,.\nonumber
\end{align}
Analogously, the $_{55}$ component of Einstein equations Eq.~\eqref{e55} reads 
\begin{equation}\label{yh3}
\ddot{Y}\left( Y+\frac{f_k(r)}{c_1}\right)=2\dot{Y}^2+kY^2+g(r)Y+h(r)\,,
\end{equation}
where
\begin{align}
g(r)&=e^{-R}\left[(kr^2-1)\left(\frac{3}{r}-\bar{R}\right)+1\right]+k\frac{f_k(r)}{c_1}\,,\nonumber\\
h(r)&=g(r)\frac{f_k(r)}{c_1}-k\frac{f_k^2(r)}{c_1^2}-2\left(\frac{1-(kr^2)^2}{c_1r}\right)-\frac{c_5}{3}\,.\nonumber
\end{align}
Eqs.~(\ref{yh1}), (\ref{yh2}) and (\ref{yh3}) can be reduced to first order EDOs. By defining a new variable 
$X=\dot{Y}$, Eqs.~(\ref{yh1}) and (\ref{yh2}) can be written as
\begin{equation}\label{reduc1}
2X\frac{dX}{dY}\left(Y+\alpha\right) = 3X^2 + kY^2 + bY + c,
\end{equation}
where $\alpha = f_k(r)/c_1$, for 
Eqs.~(\ref{yh1}) and (\ref{yh2}), by identifying respectively $b$ to $u(r)$ and $m(r)$, and $c$ to $v(r)$ and $n(r)$.
Solutions 
are provided by
\begin{equation}
\dot{Y}^2=k_1 (Y+\alpha)^3-kY^2-\frac{1}{2}bY-\frac{1}{3}k\alpha^2-\frac{1}{6}\alpha b-\frac{1}{3}c,
\end{equation}
where $k_1$ is a constant of integration. Note that when $k_1=0$ the above equation has exactly the same form of Eq.~\eqref{eqY}, thus has the same kind of solutions.

Moreover Eq.~(\ref{yh2}) can be recast 
analogously as 
\[X\frac{dX}{dY}(Y+\alpha) = 2X^2 + kY^2 + g(r)Y + h(r)\,,\] and reduced to a first order EDO in a similar way, giving
\begin{equation}
\dot{Y}^2=k_1 (Y+\alpha)^4-kY^2-\frac{2}{3}(\alpha k+g(r))Y-\frac{1}{3}(\alpha^2k+\alpha g(r)+3h(r)).
\end{equation}
In general, the above obtained equations do not exhibit analytical solutions. However when $k_1=0$ the same form of the Eq.~\eqref{eqY} is again achieved.

Given the form of the above equations, the constant parameter $k_1$ fixes the initial acceleration associated to the scale factor of the universe. One can compare it to the expected data for the dynamics of the scale factor and set $k_1$ according to the initial conditions.

\begin{figure}[H]
\begin{minipage}{14pc}
\includegraphics[width=14pc]{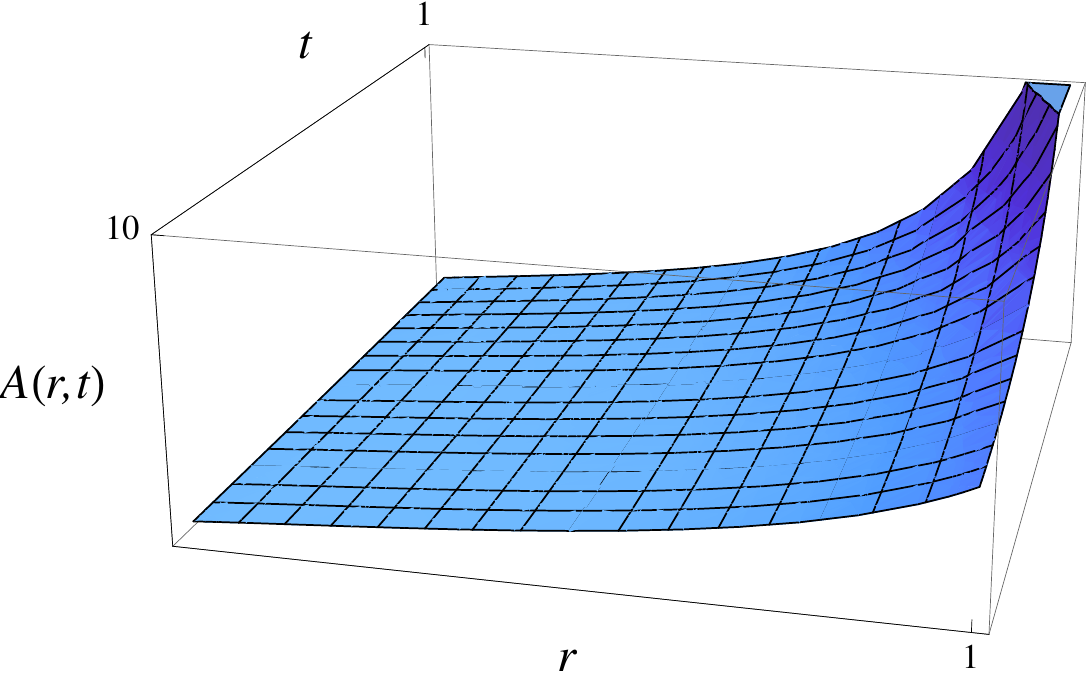}
\caption{\label{1} \footnotesize\; { Graphic of the warp factor $A_{-1}(t,r)$ in (\ref{A1}), for $c_1=2$.}}
\end{minipage}\hspace{7pc}%
\begin{minipage}{14pc}
\includegraphics[width=14pc]{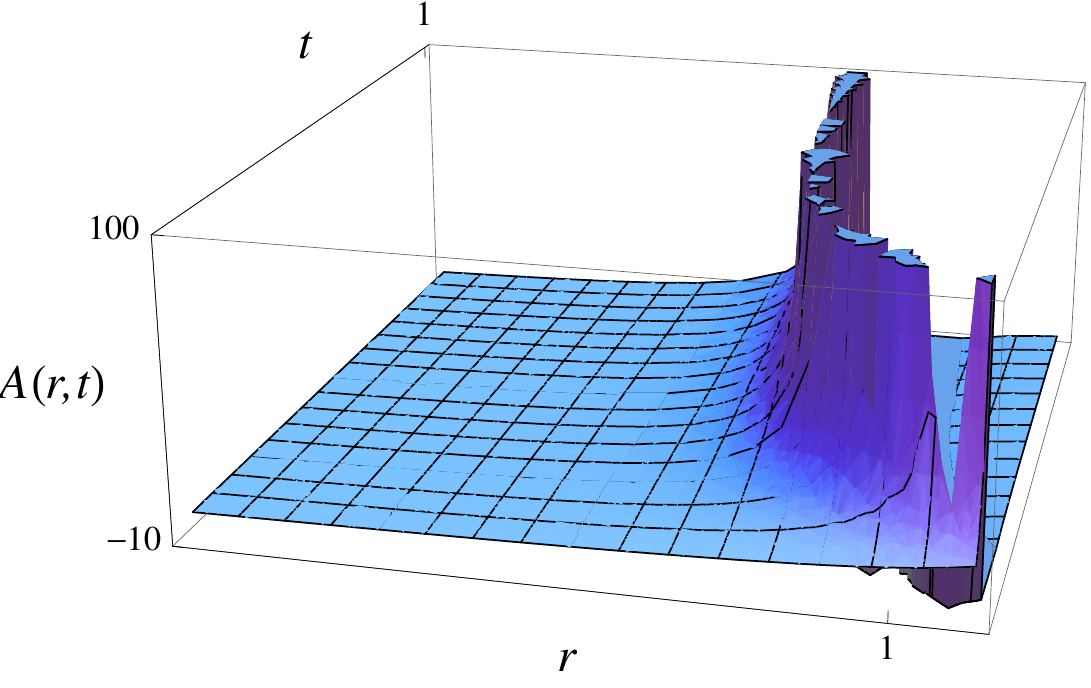}
\caption{\label{2} { \footnotesize\; Graphic of the warp factor $A_0(t,r)$ in (\ref{A1}), for $c_1=2$.
}}\end{minipage}\hspace{7pc}%
\begin{minipage}{14pc}
\includegraphics[width=14pc]{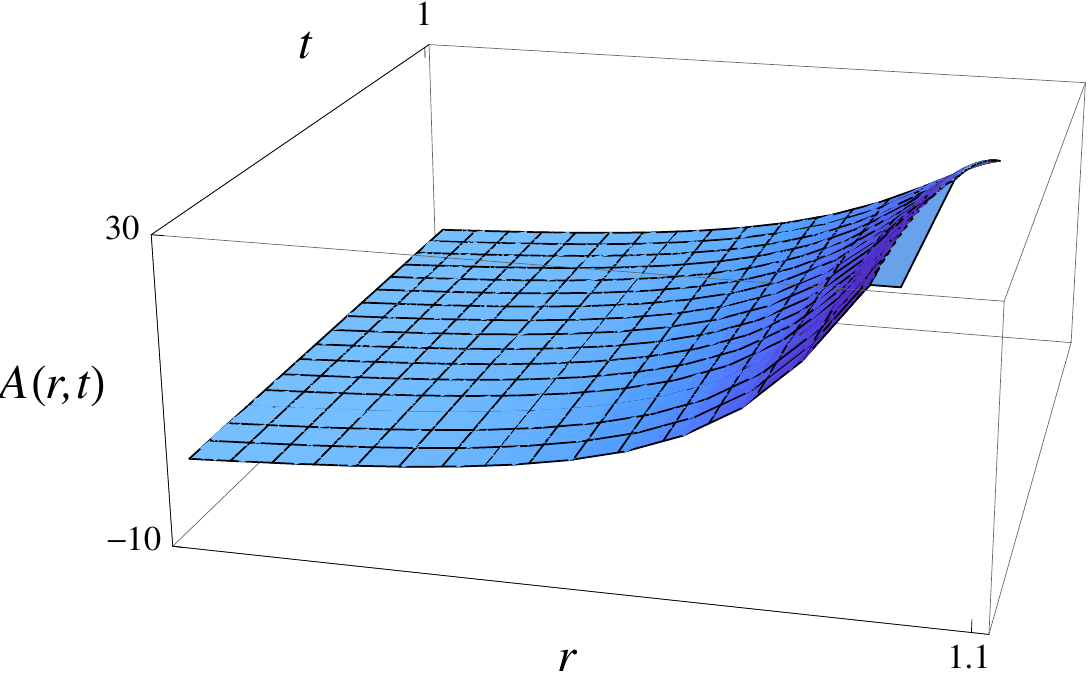}
\caption{\label{3} \footnotesize\; { Graphic of the warp factor $A_1(t,r)$ in (\ref{A1}), for $c_1=2$. }}
\end{minipage}\\
\begin{minipage}{14pc}
\includegraphics[width=14pc]{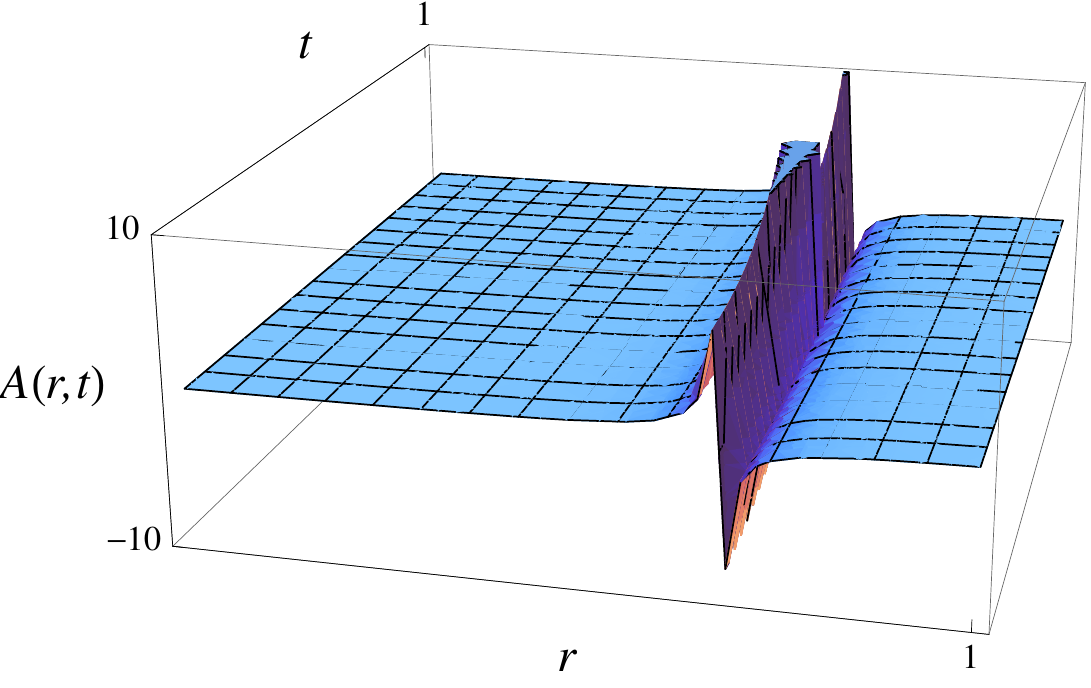}
\caption{\label{4} \footnotesize\; { Graphic of the warp factor $A_{-1}(t,r)$ in (\ref{A1}), for $c_1=0.1$. }}
\end{minipage}\hspace{7pc}%
\begin{minipage}{14pc}
\includegraphics[width=14pc]{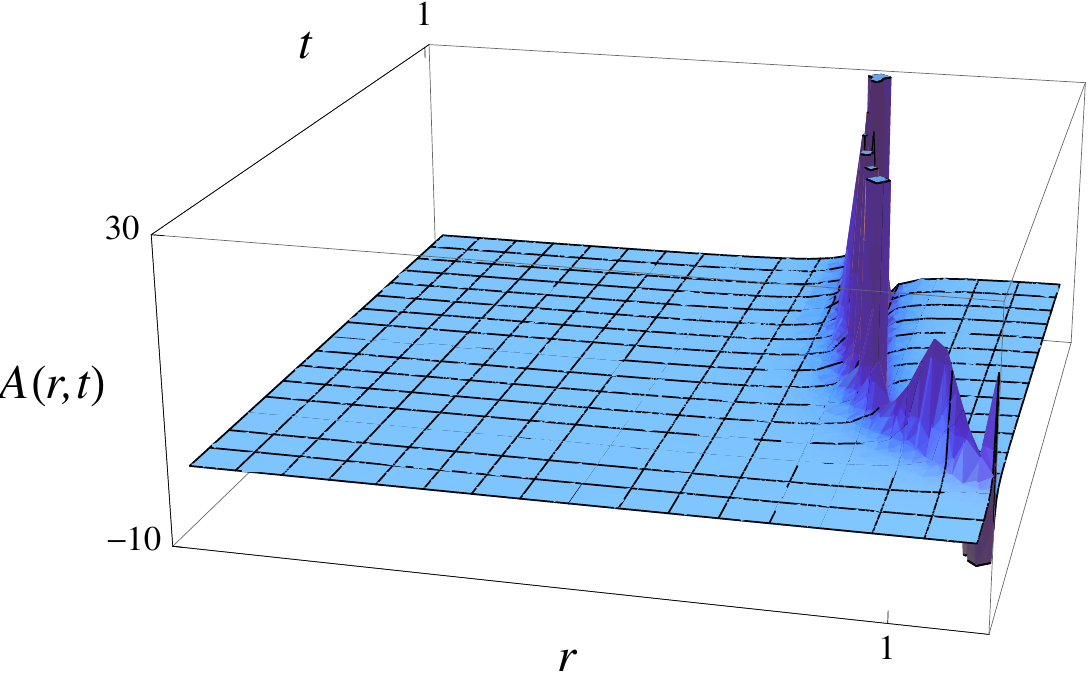}
\caption{\label{5} { \footnotesize\; Graphic of the warp factor $A_0(t,r)$ in (\ref{A1}), for $c_1=0.1$. 
}}\end{minipage}\hspace{7pc}%
\begin{minipage}{14pc}
\includegraphics[width=14pc]{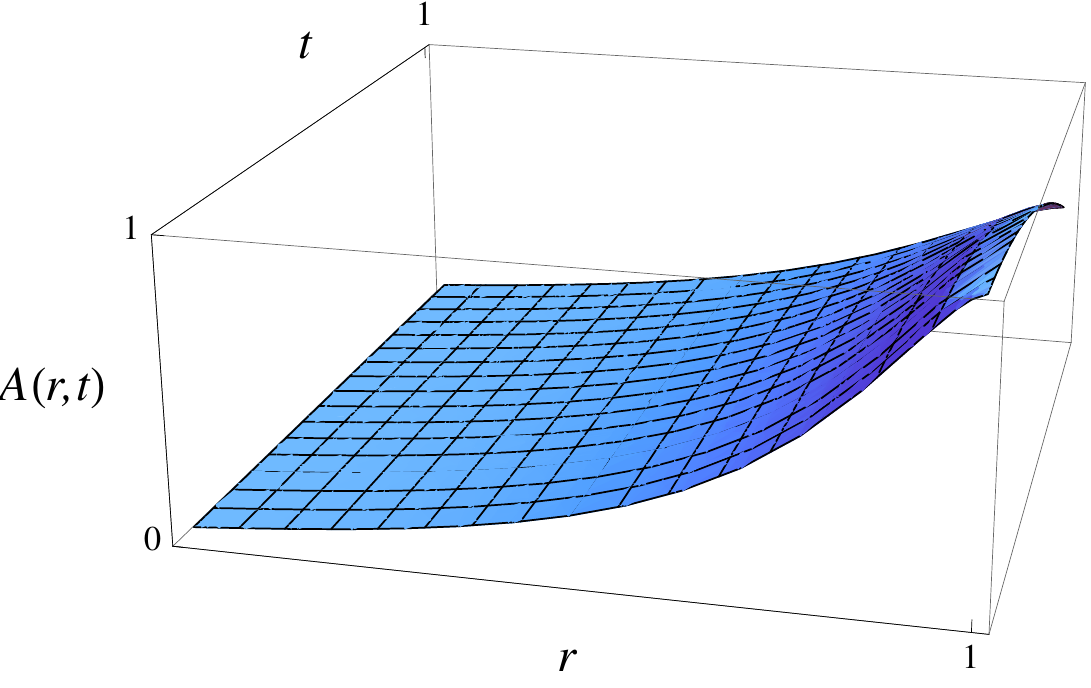}
\caption{\label{6} \footnotesize\; { Graphic of the warp factor $A_1(t,r)$ in (\ref{A1}), for $c_1=0.1$. }}
\end{minipage}
\label{figuraB}
\end{figure}

In the set of Figs.~\ref{1}-\ref{3} and Figs.~\ref{4}-\ref{6} one respectively depicts the form for the warp factor $A_k(t,r)$ for $k=0,\pm1$, and the associated Hubble like parameter, calculated from the respective warp factor.
The above two sets of plots 
indicate a dependence on the 
integration parameter $c_0$, that is a multiple of the brane cosmological constant in an isotropic thick brane-world. Instead, here 
the constant $c_0$ is related to the cosmological constant $c_5 = \Lambda$
by Eq.~(\ref{opop}), and the spherical symmetry of the warp factor sets in. This explains the dependence of the $A_k(t,r)$ on the 
parameter $c_0$.

When the constant of integration $c_1$ in (\ref{opop}) is assumed to be equal to 2, Fig.~\ref{1} illustrates a monotonically increasing scale factor both radially and temporally. The larger is the radial position on the brane the steeper is the time dependence is.
Fig.~\ref{2} presents a range of singularity that attains lower values for the radial coordinate as time elapses. Fig.~\ref{3} illustrates a scale factor that increases in the range presented therein.
However such an increment is smoother as the cosmic time elapses.
When $c_1=0.1$, Fig.~\ref{4} depicts a time-independent singularity for a fixed value $r_0 \approx 0.663$ for the scale factor of a closed Universe, whereas the singularity evinced in Fig.~\ref{2} is smoother in Fig.~\ref{5}. Finally, Fig.~\ref{6} shows a similar 
profile as that one in Fig.~\ref{3}, instead the radial increment is planer.

\begin{figure}[H]
\begin{minipage}{14pc}
\includegraphics[width=14pc]{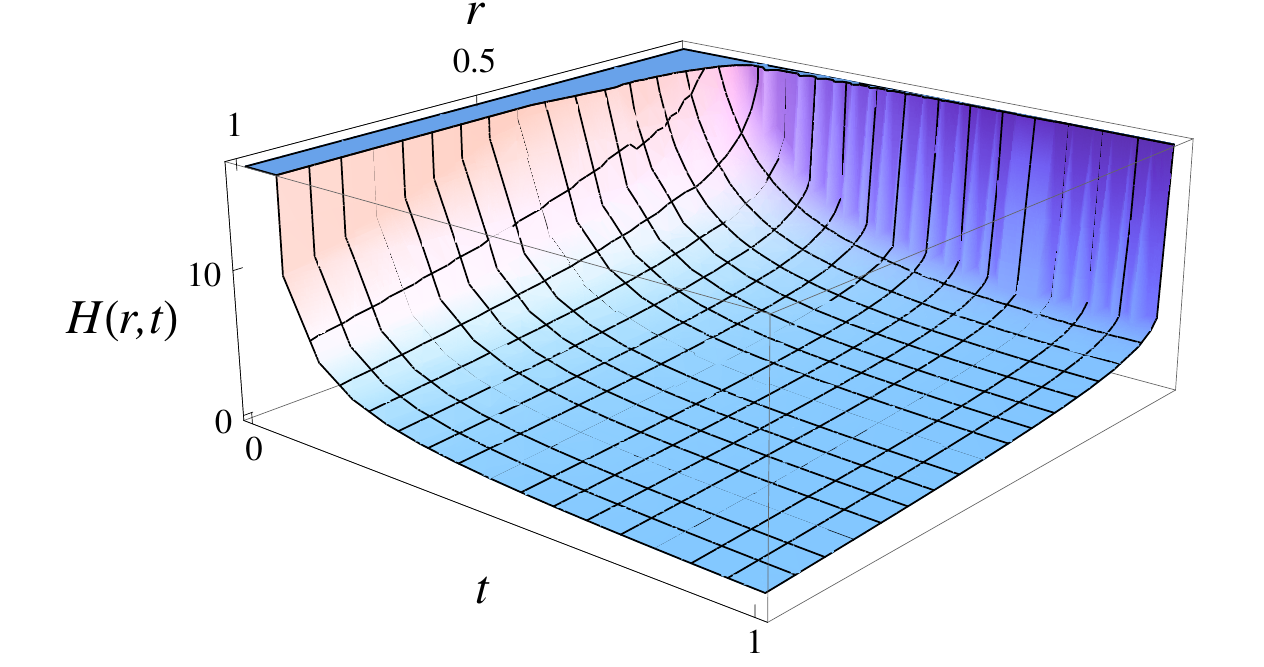}
\caption{\label{10} \footnotesize\; { Graphic of the Hubble like parameter $H_{-1}(t,r)=\dot{A}_{-1}(t,r)/A_{-1}(t,r)$ in (\ref{A1}), for $c_1=0.1$. }}
\end{minipage}\hspace{7pc}%
\begin{minipage}{14pc}
\includegraphics[width=14pc]{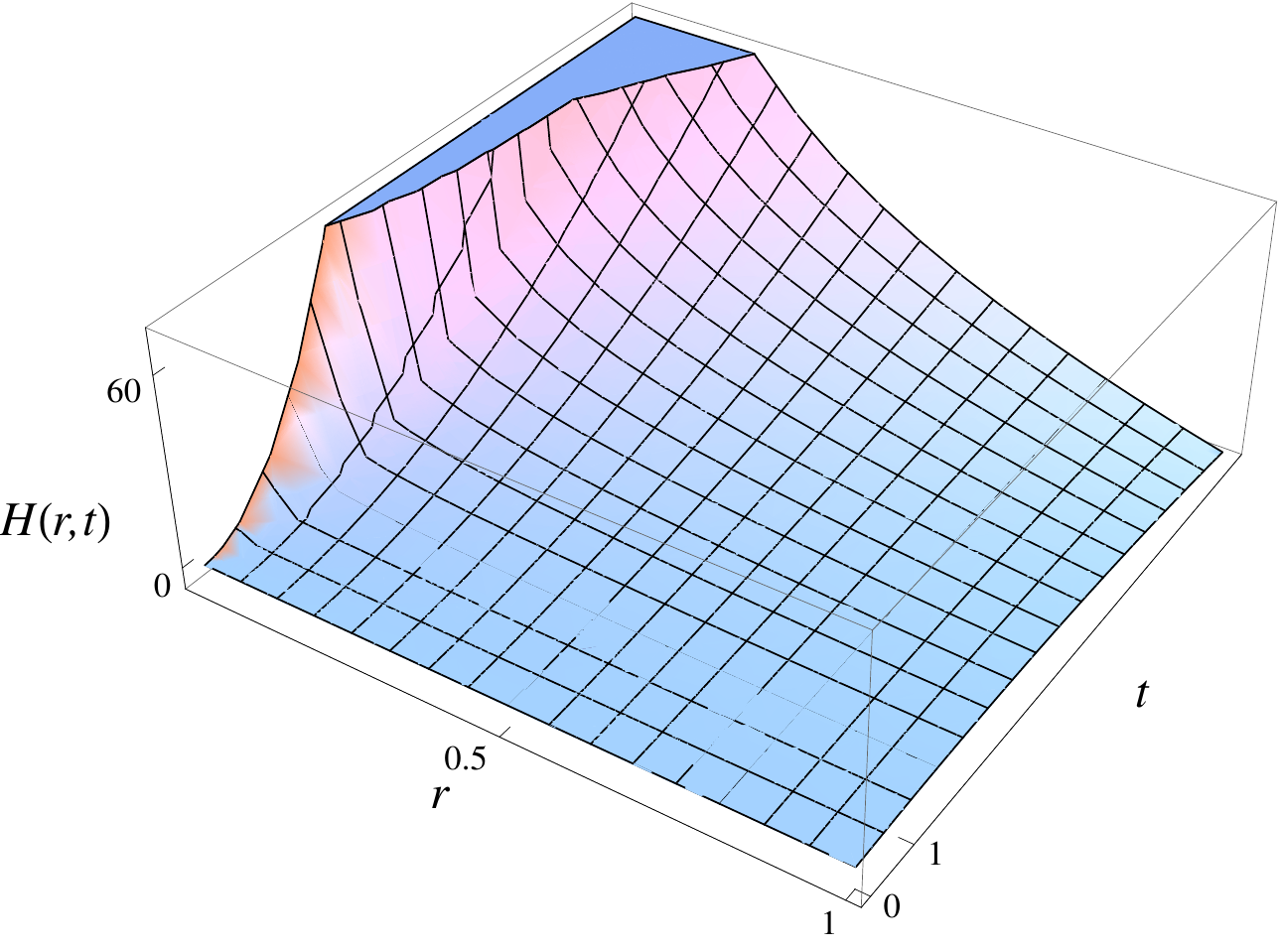}
\caption{\label{11} { \footnotesize\; Graphic of the Hubble like parameter $H_0(t,r)=\dot{A}_0(t,r)/A_0(t,r)$ in (\ref{A1}), for $c_1=0.1$. 
}}\end{minipage}\hspace{7pc}%
\begin{minipage}{14pc}
\includegraphics[width=14pc]{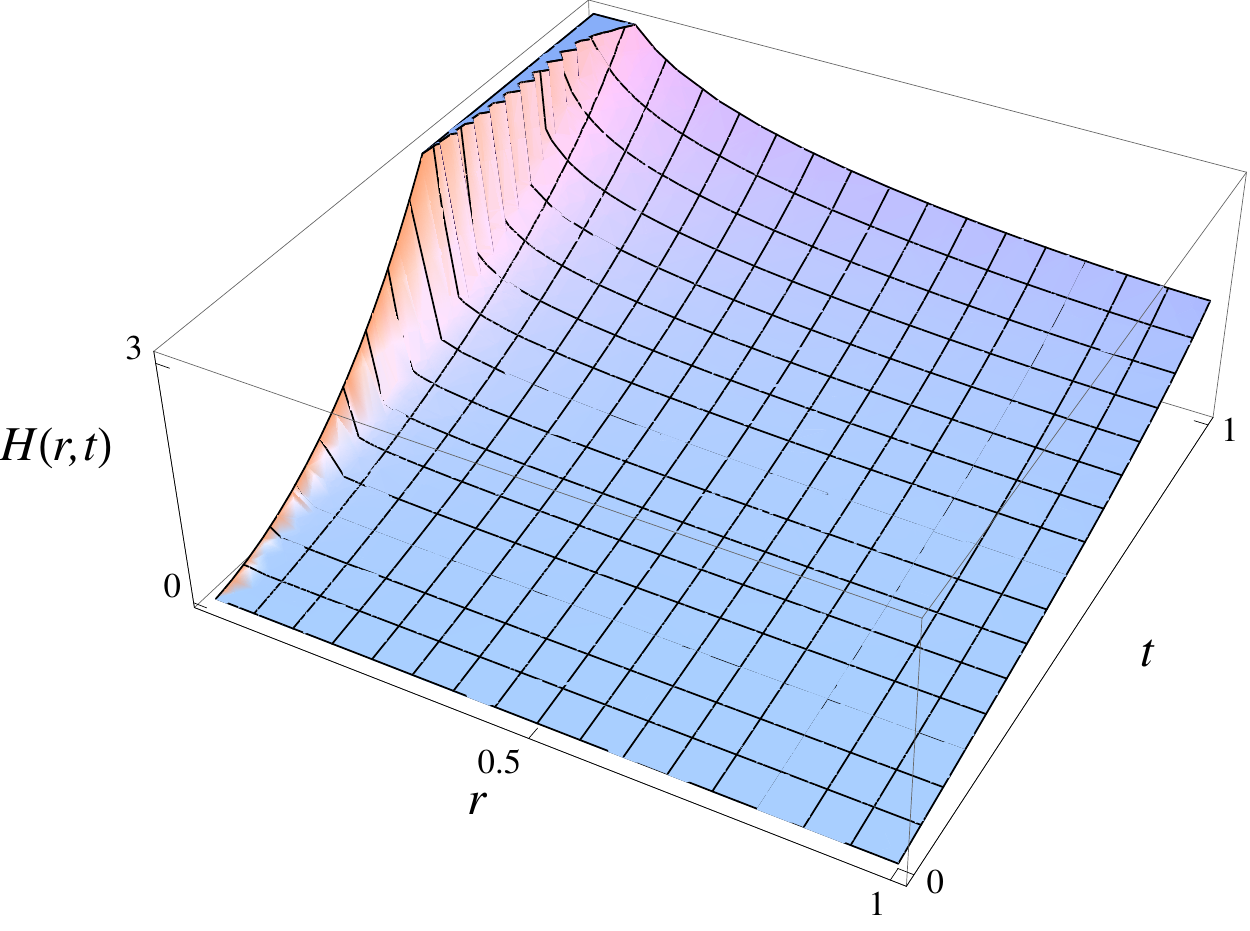}
\caption{\label{12} \footnotesize\; {Graphic of the Hubble like parameter $H_1(t,r)$ in (\ref{A1}), for $c_1=0.1$. }}
\end{minipage}
\label{figuraC}
\end{figure}

Hereupon the Hubble like parameter can also be depicted for $k=0,\pm1$. 
Their profiles are still dependent on the constant $c_1$ in (\ref{opop}), nonetheless the range of $H_k(t,r)$ changes slightly for different values of the $c_1$.
For the sake of completeness, and in order to match the results from Figs.~\ref{1}-\ref{6}, one  the Hubble like parameter for $c_1 = 0.1$ can be depicted in Figs.~\ref{10}-\ref{12}.
The above Hubble like parameters are led to the respective 
Hubble parameters  (\ref{aqeel1}), when the suitable respective limits mentioned before Eq.~(\ref{aqeel}) are taken into account. For the case $k=1$, the Hubble like parameter increases monotonically, being steepest for lower values of the radial coordinate.

The $y$-dependent part of the solutions that are determined by
Eqs.~~\eqref{by}. By defining $\bar B(y)\equiv e^{b(y)}$, it can be written as
\begin{align}
3B^{\prime\prime}+\frac32\Lambda e^{-2b}&=-\phi^{\prime2}, \label{b1}\\
6B^{\prime2}-3\Lambda e^{-2b} &=\frac{1}{2}\phi^{\prime2}-V(\phi), \label{b2}\\
\phi^{\prime\prime}+4B^{\prime}\phi^\prime -\frac{dV}{d\phi}&=0. \label{b3}
\end{align}
As such equations are the same as those ones obtained in \cite{aqeel}, our results for the extra-dimensional profiles are likewise similar here.
When one assumes a
value for the function $B(y)$, thus Eqs.~(\ref{b1})-(\ref{b3}) determine $\phi(y)$ and $V(\phi)$, or vice-versa 
\cite{wolfe,Gremm:2000dj,Afonso:2006gi,Bazeia:2012qh}. For instance, the warp factor $B(y)=\ln[\sinh(\beta y)]$ can be adopted in \cite{aqeel}, for $\beta$ usually assumed as a constant parameter, in order to have $A(y)\propto   e^{-|y|}$ when $y\to +\infty$, recovering thus the Randall-Sundrum model.
For  small (large) values of $y$, $|y|\lesssim \beta^{-1}$ ($|y|\gtrsim \beta^{-1}$), the kink solutions are given respectively by 
\ben
\phi_{\rm small}(y)&=&2\sqrt3\arctan[\tanh(\beta y/2)], \nonumber\\
\phi_{\rm large}(y)&=&\sqrt{-\frac{3\Lambda}2}\frac{1}{\beta}\sinh(\beta y),
\een

These results can be turned into asymmetric thick brane-world scenarios, generated after adding a constant to the superpotential associated to the scalar field. Asymmetric branes can be generated irrespective of the potential being symmetric or asymmetric, and the sine-Gordon-type model in this context can be shown to have a stable graviton zero mode, despite the presence of an asymmetric volcano potential \cite{Bazeia:2013usa}. 
Indeed, the superpotential method described in \cite{aqeel} can be further extended when one proposes the sine-Gordon-type model determined by the superpotential 
\[
W_c(\phi) = 2\sqrt{\frac{3}{2}}\sin\left(\sqrt{\frac{3}{2}}\phi\right)+c,
\]
that is obtained by the standard one, by shifting it by a constant parameter $c$ such that $|c|\leq\sqrt{6}$ \cite{Bazeia:2013usa}.
The 
solutions for the equations
\ben
\phi^\prime &=& \frac12 W_\phi,\label{foA}\\ 
A^\prime&=&-\frac13 W(\phi), \label{firstorderA}
\een
were obtained \cite{Bazeia:2013usa}:
\ben
\phi(y)&=& \sqrt{\frac{3}{2}}\arcsin(\tanh(y)),
\\
A_c(y)&=& -\ln[{\rm sech}(y)]-\frac13 c y,
\een 
where $\phi(y)$ is the standard solution of the sine-Gordon model, for $c=0$.
The Schr\"odinger like equation with a
quantum mechanical potential in conformal coordinates have already been studied in \cite{Bazeia:2013usa} in order to derive an stable graviton asymmetric zero model
\footnote{Other models can be still studied in this context, as for instance the twisted solutions in \cite{aqeel}, however it is out of the scope in the present paper.}.
The asymmetry induced in the thick brane-world scenario is also phenomenologically constrained to be consonant with the AdS$_5$ bulk curvature and with the experimental and theoretical limits of the brane thickness \cite{HoffdaSilva:2012em}.

The constants $c_0$ and $c_5=\Lambda$ in (\ref{c0c5}) and (\ref{e00})-(\ref{e55}) are severely constrained, besides Eq.~(\ref{opop}), by the fine tuning, relating the 4D and 5D cosmological constants, and the brane tension.
To end up, it is worthwhile to emphasize that for each $k$, the functions $w_k(r)$ and $z_k(r)$ in Eq.(\ref{eqY}) constrain the range of the variable $r$, and should be used to comply the model to observational data. The same principle must be applied in the other differential equations for $Y_k(t)$. These constraints shall be addressed in a forthcoming publication \cite{prox}. 

To finalize, further ways to analyze the system is to include the radial dependence in the bulk scalar field that supports the radial dependence in the 4D scale factor. However, in this case, it is no longer possible to solve the equations analytically.

\section*{Acknowledgements}
A. E. B. would like to thank for the financial support from the Brazilian Agencies FAPESP (grant 08/50671-0) and CNPq (grant 300809/2013-1)
R. T. C. thanks to UFABC and CAPES for financial support.
R. da R. is grateful to SISSA for the hospitality, to CNPq grants No. 473326/2013-2 and No. 303027/2012-6 and is also \emph{Bolsista da CAPES Proc. 10942/13-0.}

\end{document}